\documentclass[runningheads]{llncs}
\usepackage[T1]{fontenc}
\usepackage{graphicx}
\usepackage{hyperref}
\usepackage{color}

\usepackage{algorithm}
\usepackage{algpseudocode}
\usepackage{booktabs}
\usepackage{makecell}
\usepackage{subcaption}
\usepackage{tabularx}
\usepackage{threeparttable}
\usepackage{siunitx}
\begin{document}
\title{Taming the Search Space: Solving and Generating Hitori and Binairo Puzzles}
\titlerunning{Taming the Search Space}
%
\author{
Lukas Zandomeneghi
\and
Rainhard Dieter Findling
\and
Marc Kurz
}
\authorrunning{L.\ Zandomeneghi \and R.\ D.\ Findling \and M.\ Kurz}
%
\institute{SAIL Department, University of Applied Sciences Upper Austria}
\maketitle              
\begin{abstract}
This paper investigates solving and generation techniques for the logic puzzles Hitori and Binairo.
Two solving paradigms are compared: backtracking with domain-specific optimizations, and SAT-based solving via conjunctive normal form encodings.
An empirical evaluation analyzes runtime, explored search nodes, and branching factor across varying puzzle sizes.
To support systematic benchmarking in the evaluation, generators capable of producing valid and uniquely solvable puzzle instances are developed.
Results indicate that constraint propagation is the most effective backtracking optimization, substantially reducing the effective branching factor, search tree size, and thus runtime.
Heuristic variable ordering and scoring strategies provide additional improvements.
For Binairo, the SAT-based approach solves all evaluated instances within low runtime, while optimized backtracking fails to solve difficult puzzle instances within the timeout.
For Hitori, propagation-based backtracking achieves the best results, while for the SAT-based approach the iterative connectivity check takes up the majority of the runtime, failing difficult puzzle instances.

\end{abstract}
%
%
%

\section{Introduction}

Logic puzzles such as Hitori and Binairo combine simple rule sets with complex combinatorial structure.
Although primarily recreational, these puzzles admit formal modeling as constraint satisfaction problems (CSPs).
Efficiently solving and generating such puzzles requires navigating large search spaces under structural constraints, making them suitable benchmarks for studying algorithmic design decisions.
Two of the most effective paradigms for solving CSPs are systematic backtracking search and reduction to Boolean satisfiability (SAT).
Modern SAT solvers rely on highly optimized conflict-driven clause learning techniques, while backtracking approaches allow tight integration of domain-specific heuristics and incremental consistency checks.

Understanding the practical trade-offs between these paradigms for the Hitori and Binairo puzzles forms the central motivation of this work.
This work evaluates both SAT-based and backtracking-based solvers for Hitori and Binairo, including different backtracking heuristics and optimizations.
The SAT-based solvers encode the puzzles as propositional formulas in conjunctive normal form and delegate solving to a state-of-the-art SAT solver.
Beyond solving, dedicated generators were developed for both puzzle types to produce valid instances with unique solutions, which are used in the evaluation.
In summary, the contributions of this work are:

\begin{itemize}
    \item Backtracking optimizations and heuristics for Binairo and Hitori puzzles.
    \item SAT encodings of Binairo and Hitori puzzles.
    \item Approaches to Binairo and Hitori puzzle generation.
    \item An empirical evaluation of the impact of solvers, optimizations, and heuristics on runtime requirements, over different Binairo and Hitori puzzles sizes.
\end{itemize}

\section{Hitori, Binairo, and Constraint Satisfaction Problems}
\label{sec:background}

\begin{figure}[t]
    \centering
    \captionsetup[subfigure]{justification=centering}
    \begin{subfigure}[t]{0.24\textwidth}
        \centering
        \includegraphics[width=\linewidth]{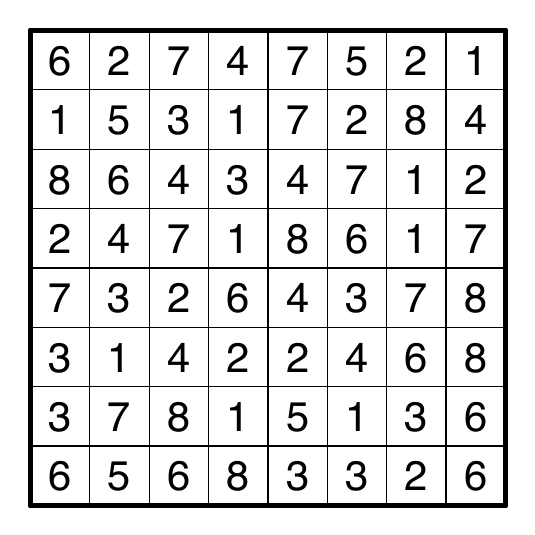}
        \caption{Unsolved Hitori}
    \end{subfigure}\hfill
    \begin{subfigure}[t]{0.24\textwidth}
        \centering
        \includegraphics[width=\linewidth]{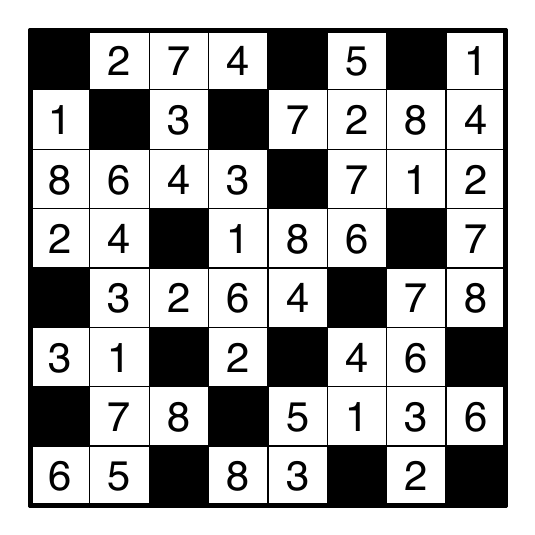}
        \caption{Solved Hitori}
    \end{subfigure}\hfill
    \begin{subfigure}[t]{0.24\textwidth}
        \centering
        \includegraphics[width=\linewidth]{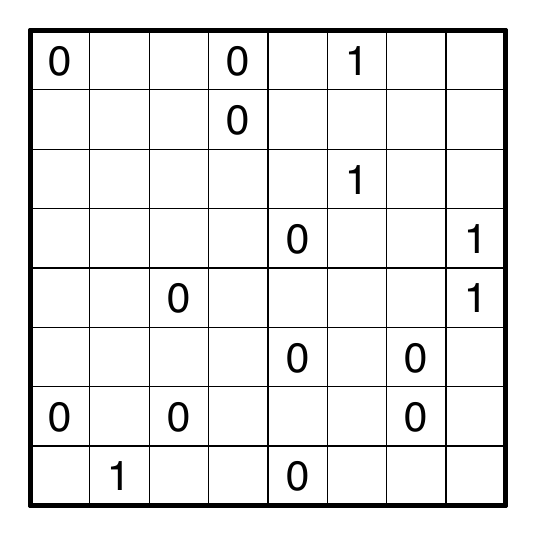}
        \caption{Unsolved Binairo}
    \end{subfigure}\hfill
    \begin{subfigure}[t]{0.24\textwidth}
        \centering
        \includegraphics[width=\linewidth]{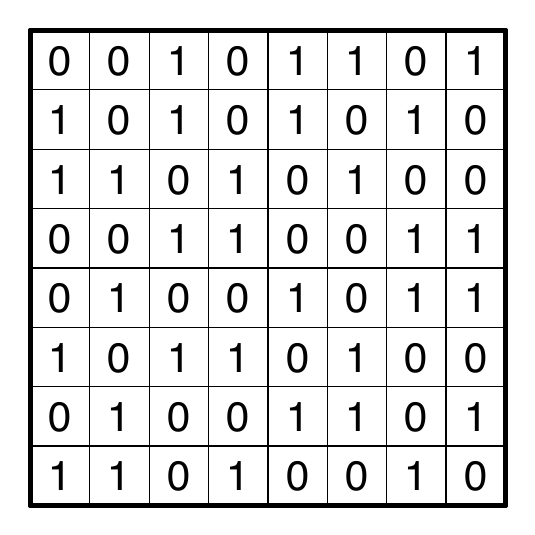}
        \caption{Solved Binairo}
    \end{subfigure}
    \caption{Example Hitori and Binairo puzzle instances before and after solving.}
    \label{fig:puzzle-examples}
\end{figure}

Hitori (Fig.~\ref{fig:puzzle-examples}) is a logic puzzle defined on an $n \times n$ grid, where each cell initially contains a natural number~\cite{pacheco2025explaining,Suzuki_2017_HitoriNumbers}. 
Initially, each cell is uncolored (often represented as gray).
The objective is to blacken a subset of cells such that the remaining cells satisfy a set of constraints.
Cells that are not blackened retain their original numbers, while black cells are considered removed from the grid.
Cells that cannot be blackened due to the constraints are considered white and retain their original numbers.
The constraints of Hitori are as follows.
First, for every row and column, no number may appear more than once among the white cells. 
Second, no two black cells may be orthogonally adjacent. 
Third, all white cells must form a single orthogonally connected component.
A solution to a Hitori puzzle is a binary coloring of the grid cells into black and white that satisfies all constraints.
A valid Hitori puzzle has exactly one valid solution.

Binairo, also called Takuzu (Fig.~\ref{fig:puzzle-examples}) is a binary logic puzzle defined on an even-sized $n \times n$ grid~\cite{Utomo_2017_Solvingbinarypuzzle}.
Initially, the grid contains some cells filled with either $0$ or $1$, while the remaining cells are empty.
The objective is to fill all empty cells with $0$ or $1$ such that the following constraints are satisfied.
First, each row and each column must contain an equal number of zeros and ones. 
Second, no row or column may contain three consecutive cells with the same value. 
Third, all rows must be pairwise distinct, and all columns must be pairwise distinct.
A solution to a Binairo puzzle is a complete assignment of 0 and 1 values to all cells that satisfies these constraints. 
A valid Binairo puzzle has exactly one valid solution.

When puzzles like Hitori and Binairo are formulated as a constraint satisfaction problem (CSP), each cell of the grid corresponds to a variable whose domain represents the set of admissible values for that cell~\cite{kumar1992algorithms,rossi2006handbook}.
For Hitori, variables range over the domain $\{\text{white}, \text{black}\}$, while the original numbers in the grid induce global constraints on rows and columns. 
For Binairo, variables range over the domain $\{0,1\}$, and the puzzle rules define constraints on local patterns, row and column counts, and global uniqueness.
The constraints in both puzzles are a mixture of local constraints, which involve neighborhoods of cells, and global constraints, which span entire rows, columns, or connected components. 
This combination leads to large search spaces with structural regularities, making both puzzles interesting for solving via backtracking search, constraint propagation, and SAT-based solving techniques.

\section{Related Work}
\label{sec:related-work}

Both Binairo and Hitori have been proven NP-complete~\cite{de2012binary,hearn2009games}, placing them in the same complexity class as Sudoku but offering distinct topological challenges, such as the global connectivity requirements found in Hitori~\cite{Suzuki_2017_HitoriNumbers}.

The foundational approach for solving such puzzles is backtracking search~\cite{van2006backtracking}.
Naive backtracking search (also called plain backtracking search) explores the state space through depth-first search (DFS), reverting decisions upon reaching a contradiction.
While easy to implement, its efficiency is highly sensitive to variable and value ordering~\cite{norvig2009sudoku}.
As an educational application of constraint satisfaction, Butler~\cite{Butler_2013_Sudokupencilpuzzles} describes a Hitori solver that uses forward checking for the duplicate and adjacency constraints and graph search for the connectivity constraint.
In the context of Binairo, Utomo and Makarim~\cite{Utomo_2017_Solvingbinarypuzzle} demonstrate that enforcing no-triples and balancing rules during search can prune significant portions of the tree, though the performance gain relative to declarative solvers remains a point of active analysis.

Optimized backtracking solvers provide improvements through the degree of look-ahead and domain reduction.
Heuristics such as Minimum Remaining Values (MRV) and Forward Checking (FC) are standard in Sudoku literature~\cite{norvig2009sudoku} but require adaptation for grid-based puzzles with different constraints.
Hitori presents a unique challenge because the non-adjacency rule for black cells and the connectivity rule for white cells expose both local and global constraints.
Berthier~\cite{Berthier_2013_Patternbasedconstraint} presents a general pattern-based framework for finite constraint satisfaction problems, illustrated primarily using Sudoku and related logic puzzles.

A significant body of work has focused on declarative modeling using SAT and Satisfiability Modulo Theories (SMT) solvers.
These approaches translate puzzle rules into propositional logic or first-order theories, leveraging conflict-driven clause learning (CDCL) to find solutions~\cite{bordeaux2006propositional,Bright_2019_Effectiveproblemsolving}.
Utomo~\cite{Utomo_2017_Solvingbinarypuzzle} provides a framework for encoding Binairo as a SAT problem, noting that SAT solvers often outperform backtracking on highly constrained instances but lack the direct interpretability of procedural search.
For Hitori, the graph-based connectivity constraint can be represented using an SMT formulation that assigns ordered labels to connected cells~\cite{Knijff_2021_Solvinggeneratingpuzzles}.

While solver analysis is common, the algorithmic generation of puzzles with unique solutions is less frequently addressed.
Suzuki et al.~\cite{Suzuki_2017_HitoriNumbers} explored the properties of Hitori Numbers, the minimum number of hints required for uniqueness, paralleling the minimum clue problem in Sudoku.
Generation typically involves a digging process: starting from a full valid grid and removing numbers until the puzzle is no longer uniquely solvable~\cite{Knijff_2021_Solvinggeneratingpuzzles}.
This process is inherently solver-dependent, as a fast solver is required to verify the uniqueness of large numbers of candidate grids during the generation pipeline.

\medskip
In summary, while prior work has focused on some puzzle types, for Hitori and Binairo there is limited investigation and systematic comparisons of baseline backtracking, backtracking enhanced with optimizations and heuristics, and SAT-based solving.
This paper addresses this gap.

\section{Solver Approaches}
\label{sec:solving-approaches}

This sections describes the SAT-based solvers, and backtracking-based solvers, optimizations, and heuristics, that the evaluation relies on.

\subsection{SAT-Based Solvers}

The SAT-based approach encodes puzzle constraints as propositional formulas in conjunctive normal form (CNF) and delegates solving to a SAT solver.
For Hitori, each cell is associated with a Boolean variable indicating whether it is black or white.
The encoding enforces duplicate constraints for rows and columns and prevents adjacent black cells.
The connectivity constraint is not directly encoded due to its complexity in CNF.
Instead, candidate solutions are checked after solving, and invalid solutions are iteratively excluded by adding blocking clauses.
For Binairo, Boolean variables represent binary cell assignments.
The encoding enforces local constraints (no three consecutive equal values), global balance constraints (equal number of zeros and ones per row and column), and uniqueness of rows and columns.
Row and column uniqueness is ensured by pairwise inequality constraints.
The SAT-based approach does not incorporate domain-specific optimizations; performance depends primarily on the quality of the encoding and the efficiency of the underlying SAT solver.

\subsection{Backtracking-Based Solvers}

The backtracking approach performs a depth-first search over partially solved puzzle instances ("partial assignments").
At each step, an unassigned cell is selected, a value is assigned, and the resulting partial assignment is checked for validity.
If a violation is detected, the branch is pruned.

The efficiency of this approach depends on optimizations and heuristics that reduce the search space or guide the search to be more efficient.
These optimizations vary in their impact and computational cost.
Some optimizations, such as propagation, directly reduce the search space by eliminating infeasible assignments, while others, such as heuristic ordering strategies, influence the order in which the search space is explored.
The optimizations listed below are evaluated in this work.
Some of those optimizations include techniques discussed in related work, such as constraint propagation, forward checking, and heuristic ordering~\cite{Berthier_2013_Patternbasedconstraint,Butler_2013_Sudokupencilpuzzles,norvig2009sudoku}.
For Binairo, the propagation rules are inspired by the no-three-consecutive and balancing constraints described by Utomo and Makarim~\cite{Utomo_2017_Solvingbinarypuzzle}.
For Hitori, the duplicate and adjacency checks, as well as the connectivity checks, are motivated by Butler's constraint-satisfaction formulation of Hitori, which uses forward checking for the first two constraints and graph search for the white-cell connectivity constraint~\cite{Butler_2013_Sudokupencilpuzzles}.

\texttt{Propagation (Hitori, Binairo):}
Deterministic constraint propagation enforces consequences of partial assignments.
Forced assignments implied by constraints are applied immediately, reducing the number of undecided variables and pruning the search space.
This can triggers short chains of additional forced assignments before the search continues.

\texttt{FastDup (Hitori):}
Duplicate constraints are enforced incrementally by maintaining occurrence counts for rows and columns.
This enables fast retrieval for next candidate values in Hitori.
The optimization avoids repeated full scans of the grid when checking whether a candidate introduces a duplicate.

\texttt{CellScore (Hitori, Binairo):}
Variable selection is guided by a scoring function that prioritizes cells expected to be highly constrained.
This heuristic aims to reduce branching by selecting variables with a higher likelihood of causing early pruning.
As a result, the solver focuses first on cells that are likely to restrict many other choices.

\texttt{WhiteFirst (Hitori):}
For Hitori, value ordering prefers assigning cells to white before black.
This reflects the structural property that valid solutions typically contain fewer black cells and aims to delay fragmentation of connected white regions.
The intent is to keep the partial solution flexible for as long as possible.

\texttt{SkipConnected (Hitori):}
Connectivity checks for Hitori are omitted when structural conditions guarantee that connectivity cannot be violated.
This reduces unnecessary overhead during search.
It is only applied in situations where the remaining unassigned cells cannot break the connected white component.

\texttt{SufficientConnected (Hitori):}
A lightweight connectivity test ensures that a valid connected configuration of white cells remains possible without requiring a full connectivity analysis.
Instead of proving connectivity exactly, it rules out partial assignments that already make a connected completion impossible.

\texttt{FastSufficientConnected (Hitori):}
An optimized connectivity check variant that restricts evaluation to regions affected by recent assignments, reducing redundant computation.
This localized update strategy avoids repeating the same connectivity reasoning after every recursive step.

\texttt{IncrementalCheck (Binairo):}
Only constraints affected by the most recent assignment are evaluated.
This avoids repeated full-grid validation.
In particular, only the relevant row, column, and nearby local constraints need to be revisited after each assignment.

\texttt{BalancedValueOrder (Binairo):}
Value ordering prefers cell assignments that maintain balance between zeros and ones in rows and columns, reducing the likelihood of early constraint violations.
The heuristic therefore steers the search away from choices that would quickly make a row or column infeasible.

\texttt{EfficientPropagation (Binairo):}
A restricted propagation variant that limits updates to regions directly affected by recent assignments, reducing overhead compared to full propagation.
This makes propagation cheaper when only a small part of the puzzle changes at each step.

\section{Generator Approaches}
\label{sec:generator-implementation}

To evaluate solver performance across different instance sizes, a broad set of puzzle instances is required.
We therefore propose generator approaches for both Hitori and Binairo and use them to produce data for our evaluation.
These approaches are designed to generate valid puzzle instances with unique solutions and use the previously introduced solvers for validity and uniqueness checks.
Randomization is included in the generation process to obtain diverse instances while preserving the rules and constraints of each puzzle type.

\subsection{Hitori Generator}

\begin{figure}[tb]
    \centering
    \includegraphics[width=\linewidth]{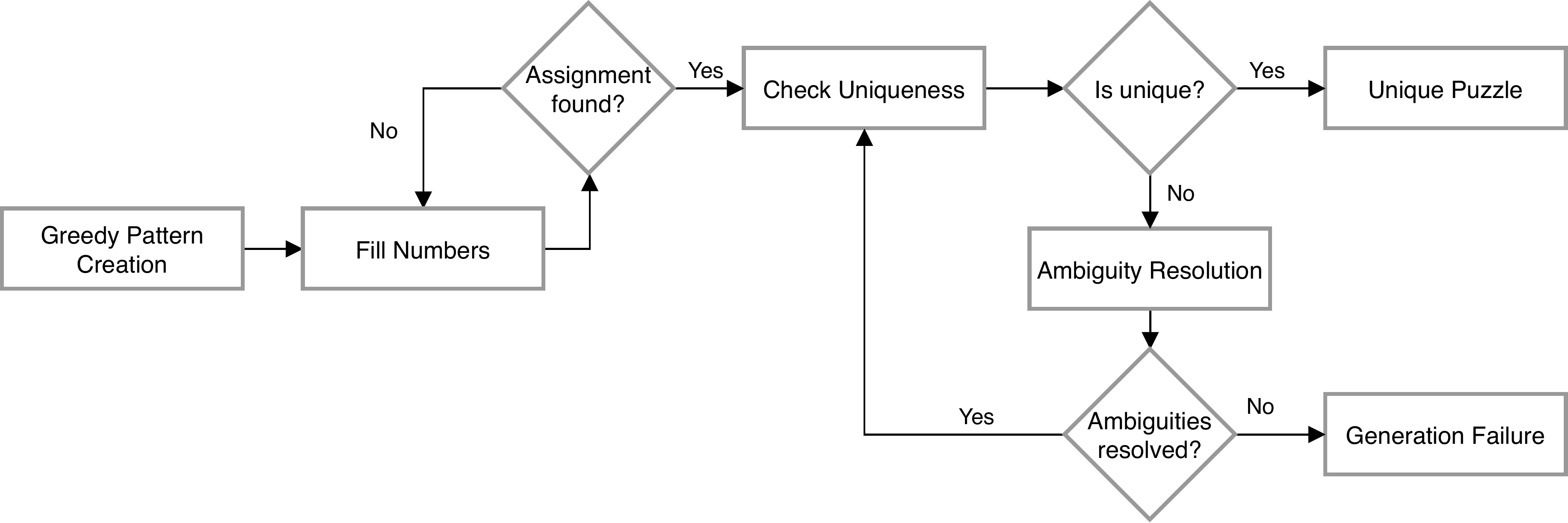}
    \caption{Generation approach of the Hitori puzzle generator.}
    \label{fig:hitori_generator_pipeline}
\end{figure}

The Hitori generator (Fig.~\ref{fig:hitori_generator_pipeline}) first constructs a valid black/white pattern by iteratively blackening random cells while enforcing Hitori constraints (no adjacent black cells, connectivity of white cells).
The white cells are then assigned numbers such that no duplicates occur within rows or columns.
The black cells are assigned numbers that appear in the same row or column, ensuring that duplicates exists.
Since this process may lead to invalid patterns, especially for smaller instances, it may be restarted until a valid configuration is obtained.

Uniqueness is verified using a modified backtracking solver that enumerates multiple solutions.
If multiple solutions are found, the generator identifies differing cells and resolves ambiguities by modifying the grid.
Small local ambiguities are handled explicitly.
In particular, if two valid solutions differ only in two cells, this ambiguity is treated as a flip pair.
The generator then modifies one of the involved rows or columns by introducing a duplicate that forces one of the two alternatives.
Ambiguities involving three differing cells are handled analogously when they can be reduced to two-cell cases.
If after resolving those ambiguities other ambiguities remain, generation is aborted.
As a result, the approach does not guarantee resolution of all ambiguities, especially for larger or non-local structures.

\subsection{Binairo Generator}

\begin{figure}[tb]
    \centering
    \includegraphics[width=\linewidth]{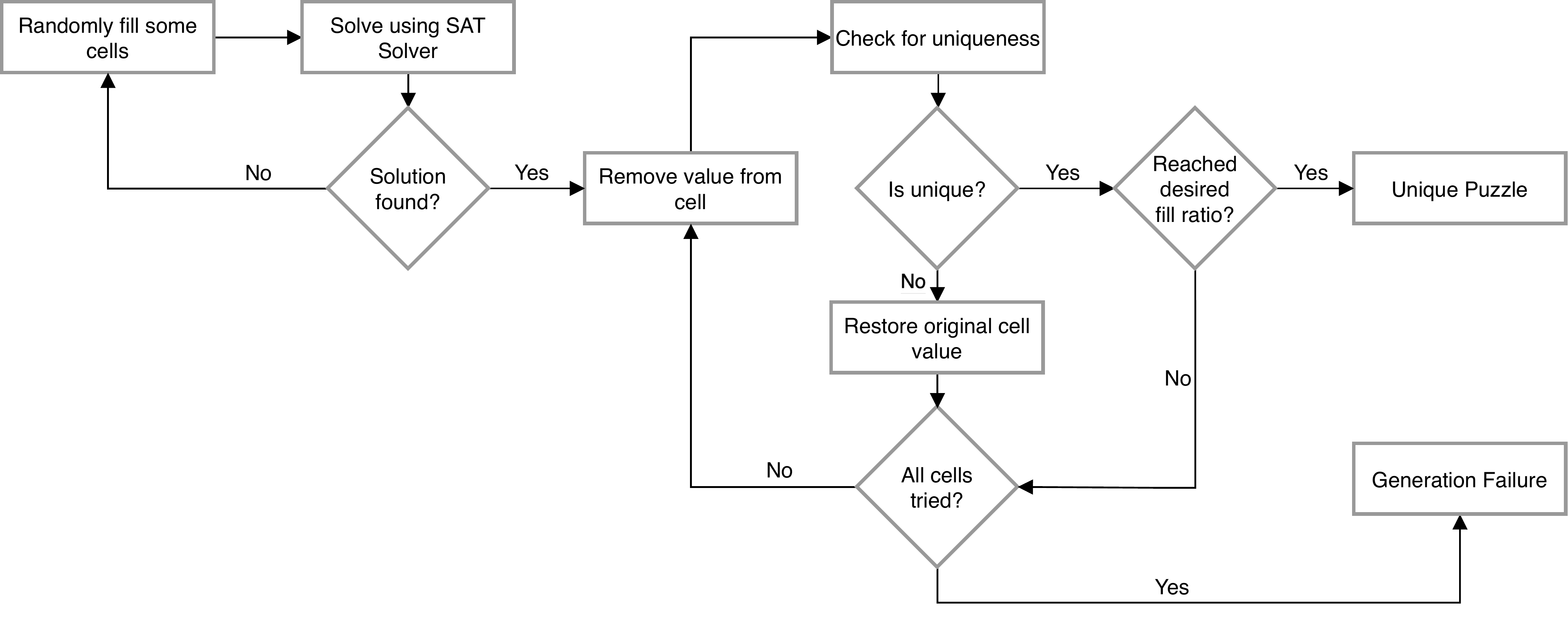}
    \caption{Generation approach of the Binairo puzzle generator.}
    \label{fig:binairo_generator_pipeline}
\end{figure}

The Binairo generator (Fig.~\ref{fig:binairo_generator_pipeline}) starts from an empty grid and randomly assigns a small fraction of cells.
This can lead to unsolvable configurations, so the process is repeated until a valid partial assignment is obtained.
By keeping the initial random assignment sparse, the chance of invalid configurations is reduced.
A SAT solver is then used to complete the grid to a full valid solution.
This avoids constructing complete valid grids through uninformed backtracking from an empty grid.

To obtain a puzzle, values are iteratively removed in random order.
After each removal, validity and uniqueness are checked using a modified backtracking solver that checks for multiple solutions.
Removals are kept only if uniqueness is preserved. This is repeated until a desired fill ratio has been reached.
The generator does not guarantee success for all instances and may terminate early if uniqueness cannot be maintained.

\section{Evaluation}
\label{sec:evaluation}

This section presents the empirical evaluation setup for the SAT-based solving and backtracking-based solving with respective optimizations.
To quantify the impact of individual backtracking optimizations and heuristics, and of their combinations, a systematic comparative study is conducted.
For both Hitori and Binairo, all possible subsets (power sets) of the available optimizations introduced in Section~\ref{sec:solving-approaches} are evaluated.
Backtracking solving with no optimizations corresponds to plain backtracking and serves as the comparison baseline.
This evaluation enables an analysis of individual optimization effects, of their combinations, and of potential redundancies among optimizations.
It is important to note that using all optimizations together does not necessarily yield the best performance, as some optimizations may introduce overhead that outweighs the benefit when combined with other optimizations.

\subsection{Test Data}

The evaluation is conducted on two benchmark sets, one for solvers and one for generators.
Each benchmark set consists solely of valid and uniquely solvable puzzle instances generated by the proposed generators.
For Binairo and Hitori, the benchmark sets include puzzle instances of grid size $8 \times 8$--$26 \times 26$, and $8 \times 8$--$54 \times 54$ respectively, both in size increments of 2.
The solver evaluation uses 100 generated instances per puzzle type and grid size.
These instances are used to evaluate and compare the selected SAT-based and backtracking-based solvers.
The generator evaluation uses a separate set of 50 generated instances per puzzle type and grid size.
This separation avoids evaluating generators on exactly the same benchmark set as used for evaluating solvers, as the former are influenced by the latter.

\subsection{Metrics and Measurement}

For each puzzle instance and solver configuration, the solver searches for a single solution and terminates upon success.
While for SAT-based solvers only runtime usage are reported, the backtracking solvers provide additional search-tree metrics characterizing the search behavior.
The following additional metrics are recorded to characterize backtracking solver behavior:
number of explored search nodes, branching factor, recursion depth, and sum of recursion depths.
Results are aggregated per puzzle grid size and solver configuration to compute the following aggregation statistics for each metric: median, 5th and 95th percentiles (P05 and P95), and the median absolute deviation (MAD).

The puzzle instance solving rate within the defined timeout (see Section~\ref{sec:experimental_setup}) is computed over all puzzle instances of a given grid size and solver configuration.
In contrast, runtime and search-tree statistics are computed only over successfully solved puzzle instances.
As a result, runtime results must be interpreted together with the respective puzzle instance solving rate.

\subsection{Experimental Setup}
\label{sec:experimental_setup}

The backtracking solvers and puzzle generators are implemented in \texttt{Python}\footnote{The solver and generator implementations are publicly available for research and teaching purposes at \url{https://codeberg.org/LukasZan/logic-puzzle-solver} and \url{https://codeberg.org/LukasZan/logic-puzzle-generators}.}. 
Numerical operations and array-based representations are primarily handled using \texttt{NumPy}. 
SAT-based solving relies on the \texttt{PySAT} framework as an interface to external SAT solvers. 
Several Solver backends supported by PySAT are evaluated during backend selection.
The evaluated SAT solvers include \texttt{cadical195}, \texttt{glucose421}, \texttt{gluecard41}, \texttt{lingeling}, \texttt{maplechrono}, \texttt{mergesat30}, \texttt{minisat22}, \texttt{maplecm}, \texttt{maplesat}, \texttt{minicard} and \texttt{minisat-gh}.

For all solver experiments, the solvers are configured to search for a single valid solution with a timeout of 5\,s.
To account for initialization overhead, one warm-up execution is performed for each puzzle instance and excluded from the reported results.
Runtime is measured as wall-clock time for each solver invocation.
All experiments are conducted on a computer with an \texttt{AMD Ryzen 9 7950X} processor and \texttt{64\,GB} of RAM, running \texttt{Windows 11}.
All solvers are executed in a single-threaded configuration to avoid variability due to parallel execution.

\section{Results and Discussion}
\label{sec:results}

\begin{figure}[tb]
    \centering
    \begin{subfigure}[t]{0.49\textwidth}
        \centering
        \includegraphics[width=\textwidth]{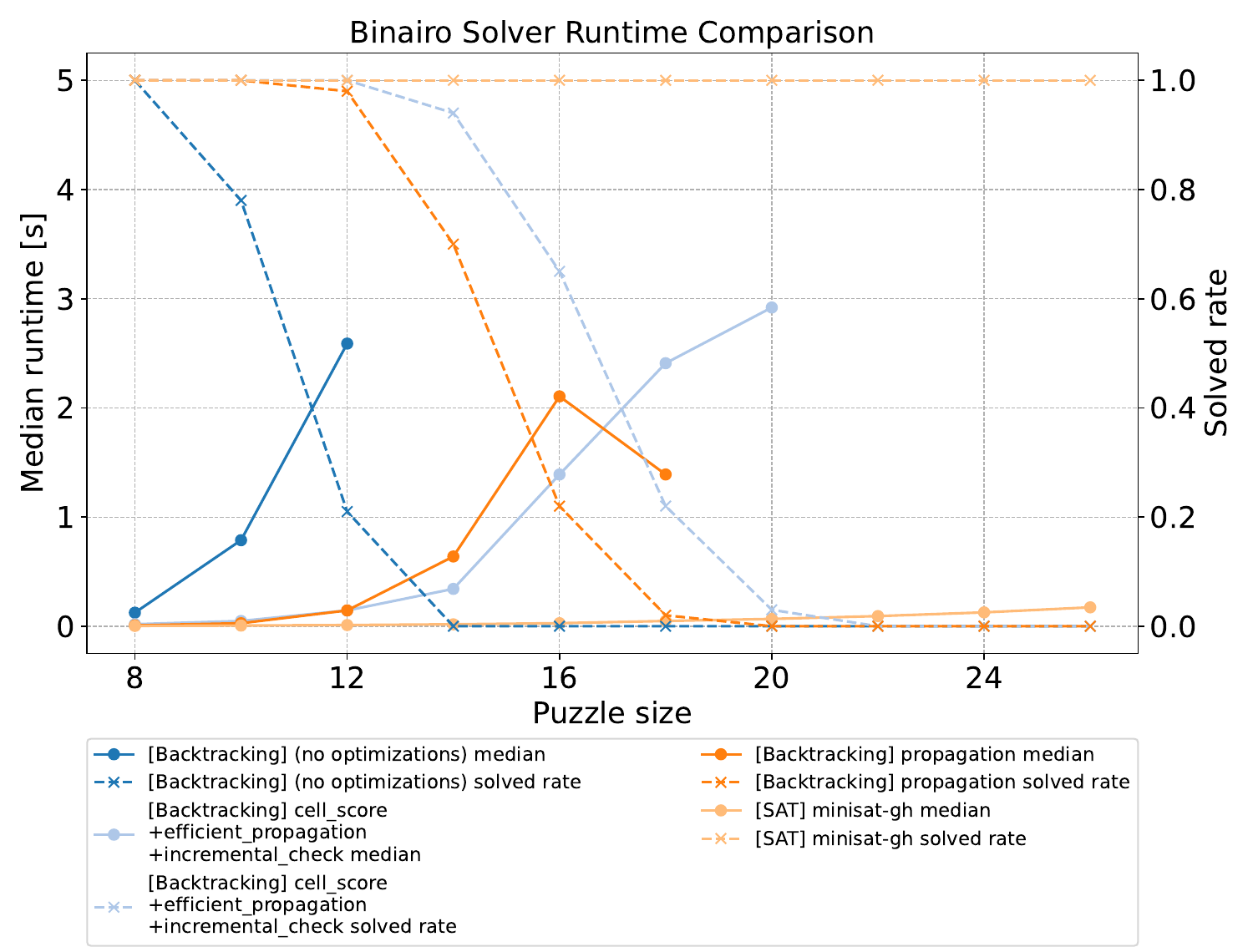}
        \caption{Binairo}
        \label{fig:binairo-solver-runtime}
    \end{subfigure}
    \hfill
    \begin{subfigure}[t]{0.49\textwidth}
        \centering
        \includegraphics[width=\textwidth]{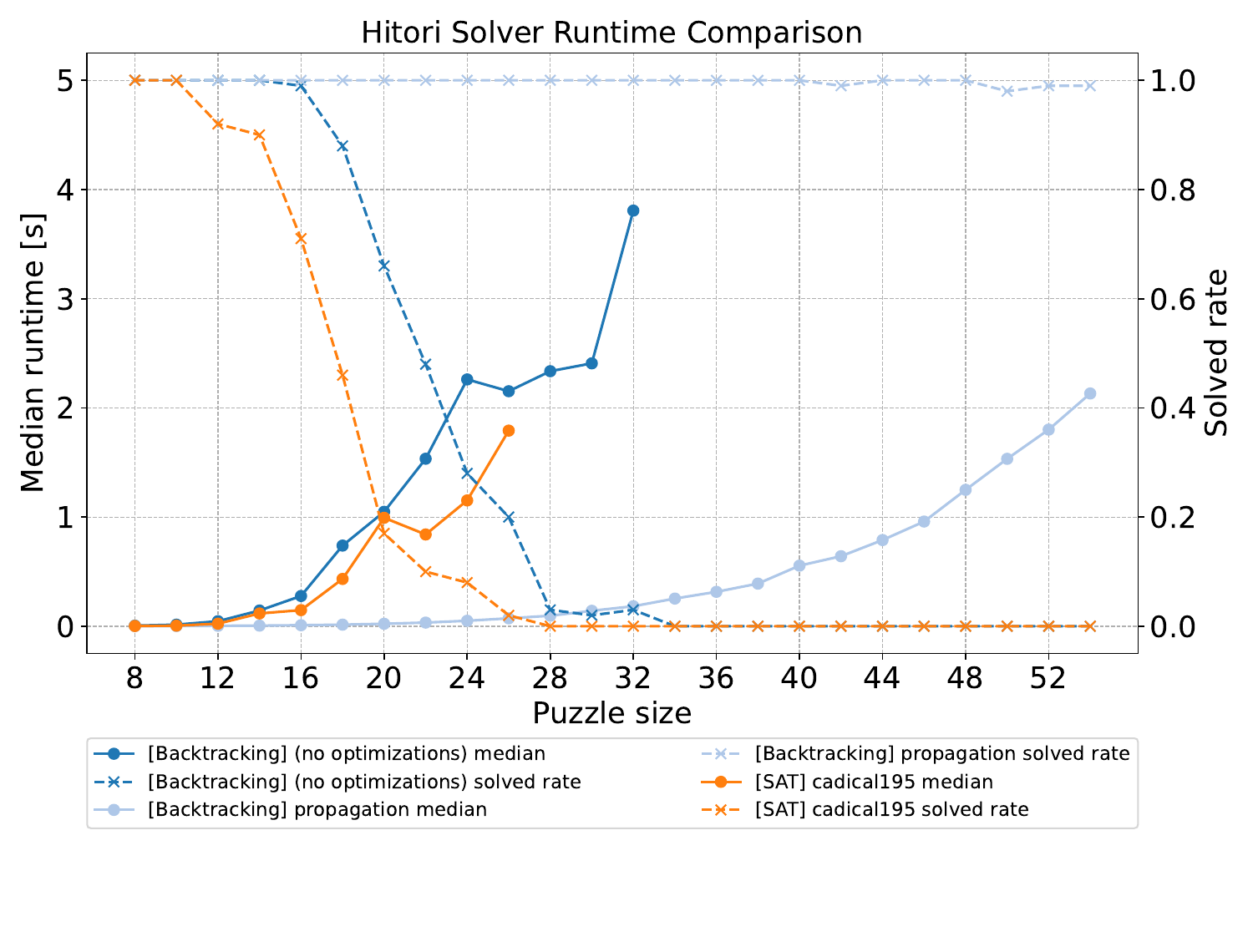}
        \caption{Hitori}
        \label{fig:hitori-solver-runtime}
    \end{subfigure}
    \caption{Runtime of selected backtracking and SAT-based solvers, for Binairo (left) and Hitori (right). Runtime statistics are computed only over successfully solved puzzle instances, while puzzle solving rate is computed over all instances.}
    \label{fig:solver-comparison-runtime}
\end{figure}

\subsection{Binairo}

Fig.~\ref{fig:binairo-solver-runtime} shows the runtime results of selected Binairo solver configurations, and Table~\ref{tab:paper-binairo-results} summarizes the corresponding aggregated results.
The SAT-based solver with \texttt{minisat-gh} yields the best results: 
It solves all puzzle instances, and achieves the lowest median runtime with $0.0383\,\mathrm{s}$ (MAD: $0.0320\,\mathrm{s}$).

\sloppy
In contrast, the best performing backtracking configuration, \texttt{CellScore+EfficientPropagation+IncrementalCheck}, only solves 48.4\% of all puzzle instances successfully within the defined timeout.
For puzzle instances solved within that timeout the median runtime is $0.1266\,\mathrm{s}$ (MAD: $0.1064\,\mathrm{s}$).
P95 runtime measurements for solved puzzle instances reveal a noticeable difference: the SAT solver P95 runtime is $0.1729\,\mathrm{s}$, while for backtracking it is noticeably higher with $2.9389\,\mathrm{s}$.
This indicates that backtracking requires noticeably more runtime specifically for difficult puzzles, which matches it not being able to solve even more difficult puzzle instances at all within the timeout.
It also indicates that the Binairo constraints are well suited for direct SAT encoding, which as a result is able to solve all puzzle instances within noticeably lower runtime.

For backtracking, \texttt{Propagation} is the most impactful optimization.
The baseline successfully solves only 19.9\% of all puzzle instances and visits a median of 4326 search nodes.
Adding \texttt{Propagation} increases the solving rate to 39.2\%, reduces the median number of visited nodes to 94, and reduces the median runtime from $0.2859\,\mathrm{s}$ (MAD: $0.2529\,\mathrm{s}$) to $0.0611\,\mathrm{s}$ (MAD: $0.0562\,\mathrm{s}$).
The result for \texttt{EfficientPropagation} indicates that the search visiting fewer search nodes does not necessarily imply lower runtime.
Although it visits fewer median nodes (58) its median runtime increases to $0.1159\,\mathrm{s}$ (MAD: $0.0940\,\mathrm{s}$).
A possible reason is that the efficient propagation process introduces additional bookkeeping and repeated checks, which regular \texttt{Propagation} does not have.
Note that the slight difference in solving rate (39.2\% to 39.1\%) is unlikely to cause such noticeable runtime increase.

\begin{table}[tb]
    \centering
    \small
    \caption{Binairo solver results for selected solvers, aggregated over all puzzle grid sizes. Runtime statistics are computed only over successfully solved puzzle instances, while puzzle solving rate is computed over all instances.}
    \label{tab:paper-binairo-results}
    \begin{tabular}{ll|r|rrr|r}
        \toprule
        Method & Configuration & Solving rate & P05 [s] & Median [s] & P95 [s] & Median nodes \\
        \midrule
        BT & None & 0.199 & 0.0183 & 0.2859 & 3.4208 & 4326.0 \\
        BT & \texttt{Propagation} & 0.392 & 0.0037 & 0.0611 & 3.0470 & 94.0 \\
        BT & \texttt{EfficientPropagation} & 0.391 & 0.0155 & 0.1159 & 2.9909 & 58.0 \\
        BT & \makecell[l]{\texttt{CellScore}\\\texttt{+EfficientPropagation}\\\texttt{+IncrementalCheck}} & 0.484 & 0.0144 & 0.1266 & 2.9389 & 49.0 \\
        SAT & \texttt{minisat-gh} & 1.000 & 0.0031 & 0.0383 & 0.1729 & -- \\
        \bottomrule
    \end{tabular}
\end{table}

\subsection{Hitori}

Fig.~\ref{fig:hitori-solver-runtime} shows the runtime results of selected Hitori solver configurations, and Table~\ref{tab:paper-hitori-results} summarizes the corresponding aggregated results.
In contrast to Binairo, \texttt{Propagation}-based backtracking yields the best results.
It solves 99.8\% of all puzzle instances, with a median runtime of $0.1594\,\mathrm{s}$ (MAD: $0.1562\,\mathrm{s}$, P95: $1.9684\,\mathrm{s}$).
The best performing SAT-based solver, \texttt{cadical195}, solves only 22.3\% of all puzzle instances within the timeout. For successfully solved puzzle instances the median runtime is noticeably lower with $0.0193\,\mathrm{s}$ (MAD: $0.0184\,\mathrm{s}$, P95: $1.8594\,\mathrm{s}$).
This indicates that the SAT solver solves easier puzzle instances successfully and noticeably faster than backtracking~-- while for harder puzzle instances it frequently exceeds the timeout, which backtracking rarely does.

Similar to Binairo, for Hitori, \texttt{Propagation} is the most impactful optimization for backtracking.
The baseline solves only 31.5\% of all puzzle instances and visits a median of 363 search nodes.
Adding \texttt{Propagation} increases the solving rate to 99.8\% and reduces the median visited search nodes to 18.
The measured median runtime increase when using \texttt{Propagation} ($0.1566$\,s to $0.1594$\,s) can be explained by the increased solving rate, as the latter includes runtime of more difficult puzzles that the baseline was not able to solve within the timeout.
This also matches the P95 runtime decrease when using \texttt{Propagation} ($3.5344$\,s to $1.9684$\,s), indicating that baseline runtime for difficult puzzle instances that are still solved within the timeout is noticeably higher, where even more difficult puzzle instances are not solved within the timeout, and hence do not contribute to the runtime measure.
Adding \texttt{SkipConnected} to \texttt{Propagation} does not substantially change runtime, solving rate, or node count.

\begin{table}[tb]
    \centering
    \small
    \caption{Hitori solver results for selected solvers, aggregated over all puzzle grid sizes. Runtime statistics are computed only over successfully solved puzzle instances, while puzzle solving rate is computed over all instances.}
    \label{tab:paper-hitori-results}
    \begin{tabular}{ll|r|rrr|r}
        \toprule
        Method & Configuration & Solving rate & P05 [s] & Median [s] & P95 [s] & Median nodes \\
        \midrule
        BT & None & 0.315 & 0.0044 & 0.1566 & 3.5344 & 363.0 \\
        BT & \texttt{Propagation} & 0.998 & 0.0014 & 0.1594 & 1.9684 & 18.0 \\
        BT & \makecell[l]{\texttt{Propagation}\\\texttt{+SkipConnected}} & 0.998 & 0.0014 & 0.1598 & 1.9571 & 19.0 \\
        SAT & \texttt{cadical195} & 0.223 & 0.0008 & 0.0193 & 1.8594 & -- \\
        \bottomrule
    \end{tabular}
\end{table}

The low puzzle instance solving rate of the SAT solver can be explained from how the white-cell connectivity constraint is handled.
Unlike duplicate and adjacency constraints, connectivity is not encoded completely in the initial SAT formula.
Instead, the solver first finds a candidate assignment satisfying the encoded constraints.
This candidate is then checked for connected white cells outside the SAT solver.
If the white cells are disconnected, a blocking clause is added and the SAT solver is called again.

The number of repeated SAT solver calls required to solve a puzzle instance is one indicator of the cost of this iterative refinement loop.
Table~\ref{tab:hitori-sat-runs} summarizes the refinement behavior for the selected SAT solver \texttt{cadical195} backend.
Those results only cover puzzle sizes for which the SAT solver solved at least one instance within the timeout.
Consequently, it covers sizes from $8{\times}8$--$26{\times}26$, as for sizes $28{\times}28$--$54{\times}54$ no puzzle instance was solved within the timeout~-- which is why the solving rate in this table differs from the solving rate in the previous Table~\ref{tab:paper-hitori-results}.

Across all sizes from $8{\times}8$--$26{\times}26$, the solving rate is 54\% with a median of 4861.5 SAT solver calls.
For sizes $8{\times}8$--$16{\times}16$, the solver achieves a solving rate of 91\%, with a median of 106.5 SAT solver calls.
For sizes $18{\times}18$--$26{\times}26$, the solving rate decreases to 17\%, while the median number of SAT solver calls increases to 9454.
These results indicate that larger and more difficult puzzle instances frequently produce disconnected candidate assignments.
Each rejected candidate requires an additional connectivity check, the generation of a blocking clause, and another invocation of the SAT solver.

\begin{table}[tb]
    \centering
    \small
    \caption{Hitori SAT refinement behavior for the selected \texttt{cadical195} backend. Only puzzle sizes for which at least one instance was solved are included. SAT solver calls indicate repeated invocations of the SAT backend before a solution was found or the timeout was reached. Solving rate and SAT solver call statistics are computed over all puzzle instances, while runtime and non-SAT share are computed only over successfully solved puzzle instances.}
    \label{tab:hitori-sat-runs}
    \begin{tabular}{l r @{\hspace{0.6em}|\hspace{0.6em}} r r r @{\hspace{0.6em}|\hspace{0.6em}} r r}
        \toprule
        Puzzle size & \makecell{Solved \\ rate} & \multicolumn{3}{c @{\hspace{0.6em}|\hspace{0.6em}}}{\makecell{SAT solver\\calls}} & \multicolumn{2}{c}{Solved puzzle instances} \\
        & & P05 & Median & P95 & Median time [s] & Non-SAT [\%] \\
        \midrule
        $8{\times}8$--$16{\times}16$ & 0.91 & 1.0 & 106.5 & 11346.3 & 0.0108 & 94.9 \\
        $18{\times}18$--$26{\times}26$ & 0.17 & 496.8 & 9454.0 & 11159.0 & 0.8560 & 96.3 \\
        \midrule
        All analyzed sizes & 0.54 & 2.0 & 4861.5 & 11238.5 & 0.0193 & 95.2 \\
        \bottomrule
    \end{tabular}
\end{table}

\begin{figure}[tb]
    \centering
    \includegraphics[width=0.75\linewidth,height=0.28\textheight,keepaspectratio]{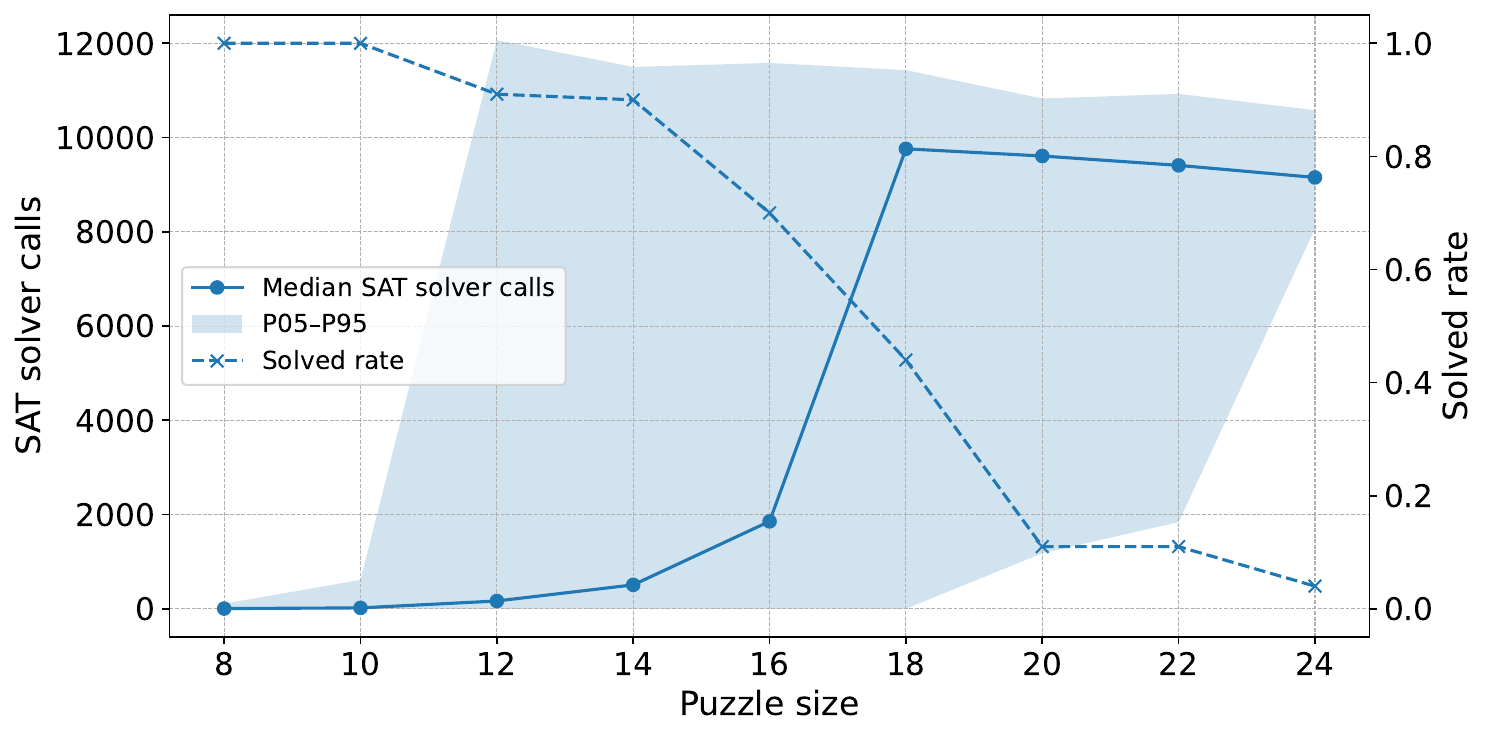}
    \caption{SAT solver calls of the Hitori SAT solver by puzzle size for the selected \texttt{cadical195} backend. The line shows the median, while the shaded area indicates the 5th to 95th percentile.}
    \label{fig:hitori-sat-runs-by-size}
\end{figure}

Fig.~\ref{fig:hitori-sat-runs-by-size} shows the number of SAT solver calls and the solving rate by puzzle size.
The median number of SAT solver calls increases substantially for the larger analyzed sizes.
The P05--P95 range also shows a high variation between puzzle instances.
Some instances terminate after only a small number of SAT solver calls, whereas others require thousands of disconnected candidate assignments to be rejected before a solution is found or the timeout is reached.
Sizes from $28{\times}28$ onward are not included in the figure because the SAT solver did not solve any instance of these sizes.

The non-SAT share of the total runtime shows that most of the runtime is spent outside the individual SAT solver calls.
For sizes $8{\times}8$--$16{\times}16$, the median non-SAT share is already 94.9\%.
For sizes $18{\times}18$--$26{\times}26$, this share increases further to 96.3\%.
The largest part of the runtime is therefore caused by the operations surrounding the SAT solver calls, particularly the global white-cell connectivity checks, the generation of blocking clauses, and the repeated coordination of the refinement loop.
This explains why the SAT-based Hitori solving approach is fast for puzzle instances that require only a small number of refinement iterations, but performs poorly when many disconnected candidate assignments must be rejected.

\medskip

Overall, those results indicate that the performance of the solving approach depends strongly on the puzzle structure.
For Binairo, SAT solving yields best results because the Binairo constraints can be encoded directly and effectively in SAT clauses.
For Hitori, the global white-cell connectivity constraint makes the evaluated SAT solving approach less suitable, as it requires an expensive iterative refinement loop that repeatedly checks for connectivity, adds blocking clauses, and reruns the SAT solver. This takes up a large part of the total runtime and frequently fails to solve the largest analyzed puzzle sizes within the timeout.
Across both puzzle types, \texttt{Propagation} is the most important backtracking optimization.
However, the Binairo result for \texttt{EfficientPropagation} indicates that reducing the number of visited nodes does not always result in less runtime, as the bookkeeping overhead required for such optimized propagation can outweigh the benefit of a smaller search tree.

\section{Conclusion and Future Work}
\label{sec:conclusion}

This paper investigated solving and generation approaches for the logic puzzles Hitori and Binairo.
Two solving paradigms were evaluated: optimized backtracking and SAT-based solving via CNF encodings.
In addition, generators for producing valid and uniquely solvable puzzle instances were developed.

The empirical evaluation shows that constraint propagation is the most effective optimization for backtracking-based solving.
Across Hitori and Binairo, propagation noticeably reduces the number of explored search nodes, thereby improving effectiveness.
For Binairo, propagation is the most effective individual optimization, while the best backtracking configuration combines cell scoring, efficient propagation, and incremental constraint checking. However, for Binairo, backtracking fails to solve difficult puzzle instances within the defined timeout.
In contrast, the SAT-based solver achieves the best overall result by solving all evaluated instances with low runtime.

For Hitori, propagation-based backtracking solves 99.8\% of all evaluated instances, whereas the SAT-based solver fails to solve difficult puzzle instances within the defined timeout. Again, propagation is the most effective optimization for backtracking-based solving.
For SAT-based solving, the low solving rate results from repeated generation and rejection of disconnected candidate solutions, which accounts for the majority of the total runtime, especially for difficult puzzle instances.
Overall this indicates that solver performance depends strongly on how well the solving paradigm matches the structure of the puzzle.

Future work could investigate more direct SAT encodings for Hitori connectivity and a more systematic use of incremental SAT solving.
Hybrid approaches may also be useful, for example, combining propagation-based preprocessing with SAT solving, or using SAT solvers for selected subproblems inside a backtracking search.
Finally, the generators developed in this paper could be extended with systematic difficulty estimation, allowing benchmark sets with different puzzle sizes and different estimated solving difficulty.

\begin{credits}

\subsubsection{\discintname}
The authors have no competing interests to declare that are relevant to the content of this paper.
Rainhard Dieter Findling is also employed at Google LLC.
\end{credits}
%
%
%
\bibliographystyle{splncs04}
\bibliography{references}

@article{bordeaux2006propositional,
  title={Propositional satisfiability and constraint programming: A comparative survey},
  author={Bordeaux, Lucas and Hamadi, Youssef and Zhang, Lintao},
  journal={ACM Computing Surveys (CSUR)},
  volume={38},
  number={4},
  pages={12--es},
  year={2006},
  publisher={ACM New York, NY, USA}
}

@article{kumar1992algorithms,
  title={Algorithms for constraint-satisfaction problems: A survey},
  author={Kumar, Vipin},
  journal={AI magazine},
  volume={13},
  number={1},
  pages={32--32},
  year={1992}
}

@incollection{van2006backtracking,
  title={Backtracking search algorithms},
  author={Van Beek, Peter},
  booktitle={Foundations of artificial intelligence},
  volume={2},
  pages={85--134},
  year={2006},
  publisher={Elsevier}
}

@book{rossi2006handbook,
  title={Handbook of constraint programming},
  author={Rossi, Francesca and Van Beek, Peter and Walsh, Toby},
  year={2006},
  publisher={Elsevier}
}

@article{pacheco2025explaining,
  title={Explaining {Hitori} Puzzles: Neurosymbolic Proof Staging for Sequential Decisions},
  author={Pacheco, Maria Leonor and Somenzi, Fabio and Srinivas, Dananjay and Trivedi, Ashutosh},
  journal={arXiv preprint arXiv:2508.14294},
  year={2025}
}

@book{hearn2009games,
  author = {Hearn, Robert A. and Demaine, Erik D.},
  title = {Games, Puzzles, and Computation},
  year = {2009},
  isbn = {1568813228},
  publisher = {A. K. Peters, Ltd.},
  address = {USA},
}

@Article{Butler_2013_Sudokupencilpuzzles,
  author    = {Butler, Zack},
  journal   = {Journal of Computing Sciences in Colleges},
  title     = {On beyond {Sudoku}: pencil puzzles across CS},
  year      = {2013},
  number    = {3},
  pages     = {21--28},
  volume    = {28},
  publisher = {Consortium for Computing Sciences in Colleges},
}

@Article{Suzuki_2017_HitoriNumbers,
  author    = {Suzuki, Akira and Kiyomi, Masashi and Otachi, Yota and Uchizawa, Kei and Uno, Takeaki},
  journal   = {Journal of Information Processing},
  title     = {{Hitori} Numbers},
  year      = {2017},
  pages     = {695--707},
  volume    = {25},
  publisher = {Information Processing Society of Japan},
}

@mastersthesis{Knijff_2021_Solvinggeneratingpuzzles,
  author  = {van der Knijff, Gerhard and Zantema, H and Geuvers, JH},
  school  = {Radboud University},
  title   = {Solving and generating puzzles with a connectivity constraint},
  year    = {2021},
  type    = {Bachelor's Thesis},
}

@misc{de2012binary,
  title={Binary puzzle is NP-complete},
  author={De Biasi, Marzio},
  year={2012},
  publisher={Technical report, ResearchGate},
}

@misc{norvig2009sudoku,
  author       = {Norvig, Peter},
  title        = {Solving Every {Sudoku} Puzzle},
  howpublished = {\url{https://norvig.com/sudoku.html}},
  note         = {Accessed: 2025-02-18},
  year         = {2009},
}

@InProceedings{Bright_2019_Effectiveproblemsolving,
  author    = {Bright, Curtis and Gerhard, Jürgen and Kotsireas, Ilias and Ganesh, Vijay},
  title     = {Effective Problem Solving Using SAT Solvers},
  booktitle = {Maple in Mathematics Education and Research},
  year      = {2020},
  publisher = {Springer International Publishing},
  address   = {Cham},
  pages     = {205--219},
}

@Article{Utomo_2017_Solvingbinarypuzzle,
  author    = {Utomo, Putranto H and Makarim, Rusydi H},
  journal   = {Mathematics in Computer Science},
  title     = {Solving a binary puzzle},
  year      = {2017},
  pages     = {515--526},
  volume    = {11},
  publisher = {Springer},
}

@article{Berthier_2013_Patternbasedconstraint,
  author       = {Denis Berthier},
  title        = {Pattern-Based Constraint Satisfaction and Logic Puzzles},
  journal      = {CoRR},
  volume       = {abs/1304.1628},
  year         = {2013},
}

\end{document}